\documentclass[a4paper,10pt]{article}
\usepackage[utf8]{inputenc}
\usepackage[english]{babel}
\usepackage{amsmath}
\usepackage{multicol}
\usepackage{graphicx}
\usepackage{color}
\usepackage{float}

\newcommand{\matr}[1]{\mathbf{#1}}
\newcommand{\fcc}{f.c.c.}
\newcommand{\pr}{Poisson's ratio}
\newcommand{\cb}{C1}
\newcommand{\cc}{C2}
\newcommand{\s}{\sigma'/\sigma}
\newcommand{\kB}{k_{\text{B}}}
\newcommand{\eps}{\varepsilon}
\newcommand{\EPS}{\boldsymbol\varepsilon}
\newcommand{\h}{\matr{h}}
\newcommand{\tS}{\matr{S}}
\newcommand{\tSa}{S_{\alpha\beta\gamma\delta}}
\newcommand{\Ninc}{N_{\text{inc}}}

\date{}

\title{%
The {\fcc} crystals of hard spheres with an array of $[001]$-nanochannel inclusions 
filled by the simplest hard sphere molecules
}

\author{J. W. Narojczyk}

\begin{document}
\maketitle

\begin{center}
\emph{
Institute of Molecular Physics, Polish Academy of Sciences, M. Smoluchowskiego 17, 60-179 Pozna\'n, Poland.}
\newline e-mail: {\tt narojczyk@ifmpan.poznan.pl}
\end{center}

\begin{abstract}
\end{abstract}
In this work the systems composed of particles intertacting with hard potential are investigated. These systems feature certain modifications to the crystal structure - selected particles are replaced with ones that differ slightly in their diameters. Such modifications, which can be thought of as ``inclusions'', concern particles located in cylindrical nanochannels, oriented in $[001]$ direction. In this study, for the first time, additional constrains have been imposed on the particles forming the inclusions. Namely, the replaced spheres have been randomly grouped into neighbouring pairs which were connected to form simple, di-atomic molecules. The results have been compared with previously investigated systems with similar inclusions but without the connections, i.e. filled only by spheres. The comparison of elastic properties between these systems is presented. It is shown that inclusions filled with dimers have different impact on the values of elastic compliances. It has been demonstrated that by introducing a small number of molecules made of spheres whose diameters differ from the rest of the particles forming the crystal, one is able to modify the hardness and shear resistance of the {\fcc} crystal without changing the {\pr} (with respect to the analogous system without additional constrains imposed on the inclusion particles).
\begin{multicols}{2}

\section{Introduction}
\label{SEC_Intro}

Auxetic~\cite{Eva1991Endeavour} materials, i.e. materials with a negative {\pr}~\cite{LanLif1986Per} (PR), have been known for over thirty years, but their practical applications to date are still rare. One of the reasons for this can be attributed to the lack of deep understanding of the nature of their extraordinary properties. First man-made material with negative {\pr} was reported in the late 1980s~\cite{Lak1987Scien_1}. In the same time the first thermodynamical model showing auxetic properties was proposed and rigorously solved~\cite{Kww1987MP,Kww1989PLA}. However, the start of the cyclic scientific conferences on auxetic and related systems was the reason why the studies on auxetic materials gained pace. New man-made auxetic materials in the form of foams~\cite{AllHewNew2017PSSB,CheScaPan2019PSSB,DunCleEss2019PSSB}, polymers~\cite{AldNazAld2016PSSB,VerHeGri2020PSSB} and composites~\cite{NovVesKen2020PSSB}, as well as in the specifically engineered structures~\cite{WanSheLia2017PSSB,RugHaLak2019PSSB,FarGatGri2019PSSB,LiHaLak2020PSSB,PhoAvrSil2021ApS,BaoRenQi2022PSSB,SunZhoZha2022PSSB}, nanostructures~\cite{MalPucRiz2020NanoMat}, or metamaterials~\cite{LiYinDon2018PSSBRRL,UstScaTur2021SMS,DudGatDud2021Mat,ValLak2021PSSB,AapKwwGri2016COMPb,MizGriGat2019PSSB,DudMarUll2022AdvMat,VerWagGri2022PSSB} have been reported. Novel fabrics with auxetic properties~\cite{ZulHu2019PSSB,JiaHu2019PSSB,ZulHuaHu2020PSSB,TahZhaHu2022PSSB}, targeted for personal applications have been developed. Alongside the experimental studies, new theoretical models exhibiting negative {\pr}~\cite{HooHoo2005PSSB,KvtKww2005JCP,KarPasDys2019PSSB,WanLaiShe2019PSSB,Lim2019PSSB,Lim2020PSSB,CzaLuk2020PSSB,WanKoRen2020PSSB,WanLaiRed2020PSSB,GorVolLis2021PSSB,KorZhoGal2022PSSRRL,KvtKww2022PSSRRL,Lim2022PSSB,LakHueGoy2022PSSB,GriAttVel2022PSSB} have been proposed. The search for the negative {\pr} in known and naturally occurring materials revealed that around 70\% of metals with cubic symmetry have such properties~\cite{BauShaZak1998Nature}. The latter can be observed only in the vicinity of specific crystallographic directions, thus making these metals \emph{partially} auxetic~\cite{BraHeyKww2011PSSB}. Baughman \emph{et all.} described the mechanism that explained the origins of auxetic properties in cubic monocrystals. Nevertheless, the understanding of auxeticity in the broader context (e.g. in two component materials) requires further research.

For this reason, with the help of numerical methods, studies of simple atomic models, composed of particles interacting with the purely geometric potential are performed. The subject of investigations are crystals of cubic symmetry, into which the inclusions of particles with different dimensions are introduced. Earlier research of such models in which the inclusions that simulate the introduction of another material into the crystal, in the form of axially symmetric channel $[001]$~\cite{JwnKwwKvt2019PSSB}, can strongly enhance auxetic properties of the {\fcc} hard sphere crystal. The present study aims to further these investigations by adding another level of constrains into the system. Namely, the particles forming inclusions are connected into simple di-atomic rigid molecules. The impact of these additional constrains on the elastic properties and the {\pr} is investigated.

\section{The Model}
\label{SEC_Model}
In this work models of hard particles are considered. This implies that particles interact with purely geometrical potential of the form:
\begin{equation}
\phi_{ij} =
  \begin{cases}
     \infty\hfill,& r_{ij}<\sigma_{ij},\\
     0\hfill,& r_{ij}\geq \sigma_{ij},
  \end{cases}
\label{EQ_HS}
\end{equation}
where $r_{ij}$ is the distance between the centres of particles $i$ and $j$, $\sigma_{ij}=(\sigma_i+\sigma_j)/2$, whereas $\sigma_i$, and $\sigma_j$ are the diameters of spheres $i$ and $j$. The \emph{hard sphere} (HS) potential is very simple, yet not trivial interaction, as it actually represents the short-range correlations that originate from the excluded volume effects~\cite{KwwKvtMko2003PRE,Fre2015NatureMat,KvtKww2020PSSBRRL}. Being one of the fundamental interactions used in the condensed matter physics~\cite{Fre2015NatureMat,HanMcD2006AP}, it is also the simplest one that can be used to study auxetic properties, as its thermodynamically stable crystalline phase, the {\fcc} hard sphere crystal, is partially auxetic~\cite{BraHeyKww2011PSSB}.

\begin{figure}[H]
  \includegraphics[width=0.475\textwidth]{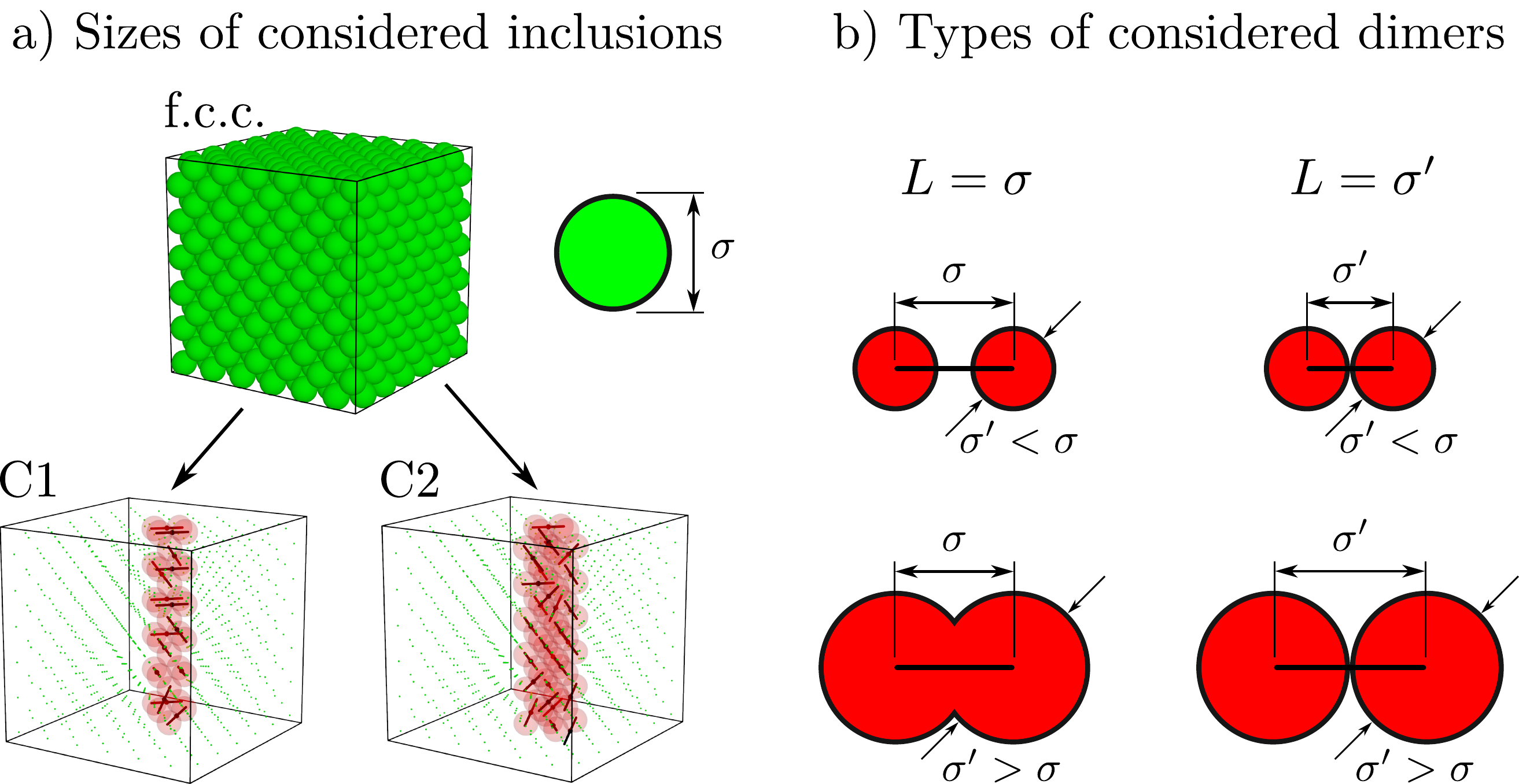}
  \caption{Details of studied systems: a) nanochannels with one (C1) and two (C2) coordination zones of spheres around selected inclusion axis, and b) the two considered cases of dimer geometry.}
  \label{FG_CH_and_MOL}
\end{figure}

The models considered in this work are modified {\fcc} crystals of $N$ hard spheres. The modification consists of a replacement of certain spheres, whose centres fall into a selected cylindrical volume, designated around selected crystalline axis, parallel to $[001]$ direction. The replaced spheres, further referred to as the \emph{inclusion} spheres, differ from the remaining ones in the value of their diameters. Whereas the regular spheres forming the crystal have the diameter equal to $\sigma$, the inclusion spheres have the diameter equal to $\sigma'$ which can be smaller or greater than $\sigma$. The primary (non-inclusion) spheres will be referred to as the \emph{matrix} spheres. The number of inclusion spheres, $\Ninc<N$, changes with the radius of the selected cylindrical volume. As the latter increases from zero (the case when only spheres lying directly on the inclusion axis are replaced) the consecutive sets of particles are gradually added to the inclusion. These sets are referred to as consecutive coordination zones of particles around the inclusion axis. The larger the number of coordination zones, the wider the nanochannel. The ratio $c=\Ninc/N$ is called the \emph{concentration}. The changes of the elastic properties will be studied with respect to different values of inclusion sphere diameters, thus the \emph{scaling factor} $s=\sigma'/\sigma$ is introduced. The systems are studied under different thermodynamic conditions with respect to different values of the dimensionless pressure $p^*=\beta\sigma^3p$, where $\beta=1/(\kB T)$, $T$  is the temperature, and $\kB$ is the Boltzmann constant.. Periodic boundary conditions are used in this study, thus effectively one obtains systems that contain arrays of parallel inclusions oriented in $[001]$ direction. In contrast to studies performed in~\cite{JwnKwwKvt2019PSSB,JwnKvtKww2022PSSB}, where such models have been investigated, in this study additional constrains have been imposed on the system. Namely, the spheres forming nanochannel inclusions have been \emph{randomly} connected in pairs, forming the simplest, di-atomic molecules, further referred to as \emph{the dimers}. The latter are considered rigid - the distance between the centres of binded spheres (the length $L$ od the dimer) cannot change. Due to the random binding of inclusion spheres, the orientation of dimers within the crystalline lattice is also random. Two cases are considered: (i) when the length $L$ of the dimer is equal to $\sigma$ (regardless of the value $\sigma'$), and (ii) when $L=\sigma'$. Visualization of studied systems is presented in Figure~\ref{FG_CH_and_MOL}.

\section{The Method}
\label{SEC_Method}

The calculations of elastic properties for the described models have been carried out with the help of computer simulations entrenched on the concept by Parrinello and Rahman~\cite{ParRah1981JAP,ParRah1982JCP}. This approach has been implemented using Monte Carlo (MC) method in the isobaric-isothermal ensemble $(NpT)$~\cite{KwwBra1989PRA,KwwKvtMko2003PRE}, i.e. for a fixed number of particles, and under constant pressure and temperature. The choice of the method was dictated by the possibility to calculate the complete elastic compliance tensor $\tS$. All the $21$ independent elements of the symmetric, fourth-rank tensor $\tSa$ are obtained from the shape fluctuations of the periodic box. The latter are directly related to the strain tensor $\eps$ - a second-rank, symmetric tensor defined as~\cite{ParRah1982JCP,KwwKvtMko2003PRE}:
\begin{equation}
 \EPS = \frac{1}{2}
  \left(\h^{-1}_p.\h.\h.\h^{-1}_p - \matr{I}\right)\ .
 \label{EQ_epsilon}
\end{equation}
In the above, $\matr{I}$ is the unit matrix of the dimensionality three and $\h_p\equiv\langle\h\rangle$ is the reference box matrix, i.e., the average box matrix at equilibrium under the temperature $T$ and the pressure $p$ or, in other words under the dimensionless pressure $p^*$. The fact that the system has the chance to optimize its shape under arbitrary applied thermodynamic conditions is one of the advantages of this approach. It is worth noting that the symmetry of the box matrix allows one to avoid deformations that would result in rotation of the system during simulations, which are unwanted in to the calculation of elastic properties. An expression describing the relation between elastic strain tensor and the elastic compliance tensor has the following form~\cite{KwwKvtMko2003PRE}:
\begin{equation}
  \tSa = \beta V_p
    \left\langle \Delta\eps_{\alpha\beta} \Delta\eps_{\gamma\delta} \right\rangle\ ,
  \label{EQ_Sijkl}
\end{equation}
where $V_p=|\text{det}(\h_p)|$ is the average volume of the system at equilibrium, under the dimensionless pressure $p^*$, $\Delta\eps_{\alpha\beta}=\eps_{\alpha\beta}-\langle\eps_{\alpha\beta}\rangle$, and $\langle\eps_{\alpha\beta}\rangle$ is the ensemble average. The Greek indices $\alpha,\beta,\gamma,\delta$ indicate directions $x$, $y$, $z$ in Cartesian coordinate system.  The general relation between $\tS$ and the {\pr} can be expressed in the following form~\cite{Tok2005PSSB}:
\begin{equation}
  \nu_{nm} =  - \frac{m_\alpha m_\beta S_{\alpha\beta\gamma\delta} n_\gamma n_\delta}
    {n_\zeta n_\eta S_{\zeta\eta\kappa\lambda} n_\kappa n_\lambda}\ .
  \label{EQ_TokNu}
\end{equation}

Elastic properties of the considered models have been determined numerically, with the use of MC simulations, carried out in the $NpT$ ensemble (within the regime of constant temperature and pressure). To study different thermodynamic conditions, the pressure has been changed. The models were subjected to four different values of external (dimensionless) pressure $p^*= 50$, $100$, and $250$. To increase the efficiency of simulations, the lowest pressure has been selected such as to avoid diffusion of particles within the crystal structure. The size of the simulated samples was $N=864$, which corresponds to $6\times 6\times 6$ {\fcc} unit cells. The number of inclusion particles varied between $30$ and $54$, depending on the size of the nanochannel. Elastic properties for each model, and each value of $p^*$ were determined for different diameters of inclusion spheres. The scaling factor $\s$ was set from the ranges of $0.95$ to $1.06$. The results were averaged over fifty independent runs for each set of parameters. The following section discusses only the results obtained for stable tetragonal systems. Thus, the presented range of $\s$ for individual models and thermodynamic conditions differ. Each simulation run took $10^7$ MC cycles. The first $10^6$ of which was treated as the period when the system reached thermodynamic equilibrium, and has been removed from calculations of averages. One Monte Carlo cycle is understood as a period in which an attempt has been made to move each of the particles once. Due to the fact that the considered system is a mixture of spheres and non-spherically symmetric dimers, independent of each translational trial move, a trial rotation of their longitudinal orientation (by small, random angles) is attempted. The acceptance ratio for translational and rotational trial moves (calculated separately) is kept at $40\%$.

\section{Results and Discussion}
\label{SEC_RandD}

\begin{figure*}
  \includegraphics[width=\textwidth]{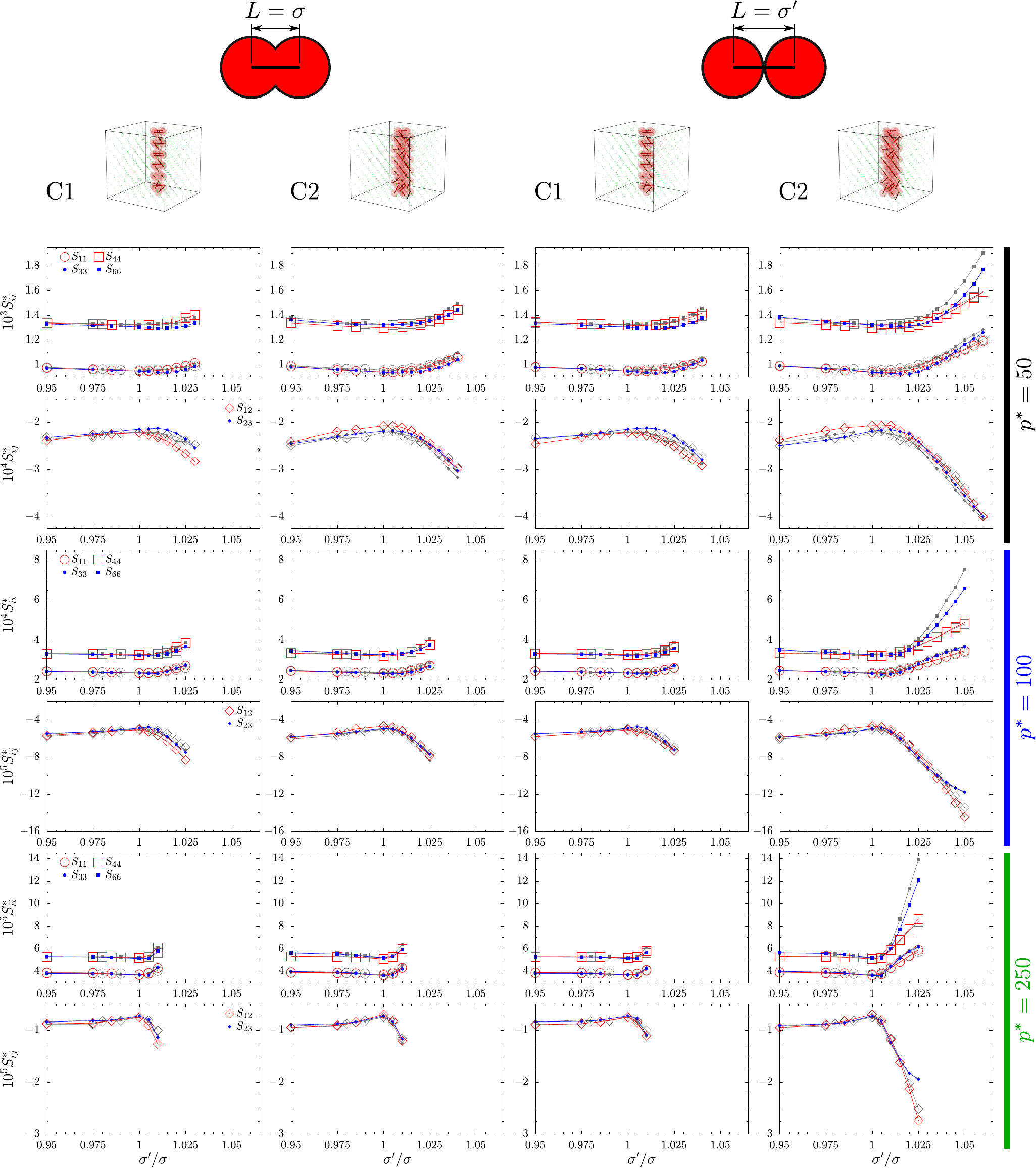}
  \caption{Elastic compliances $S_{ij}$ of tetragonal system with $[001]$-nanochannel inclusions filled with hard dimers (in colour) compared with compliances of systems containing $[001]$-nanochannel inclusions filled with hard spheres (in gray). The pressure values for the corresponding data sets have been marked with colour labels on the right. The data has been arranged in columns for systems where dimers had length equal to $\sigma$ (two columns on the left) and for systems where dimers had length equal to $\sigma'$ (right two columns). The columns correspond to the given width of nanochannels, indicated at the top with C$x$ markers.}
  \label{FG_SijBenchmark}
\end{figure*}

\begin{figure*}
  \includegraphics[width=\textwidth]{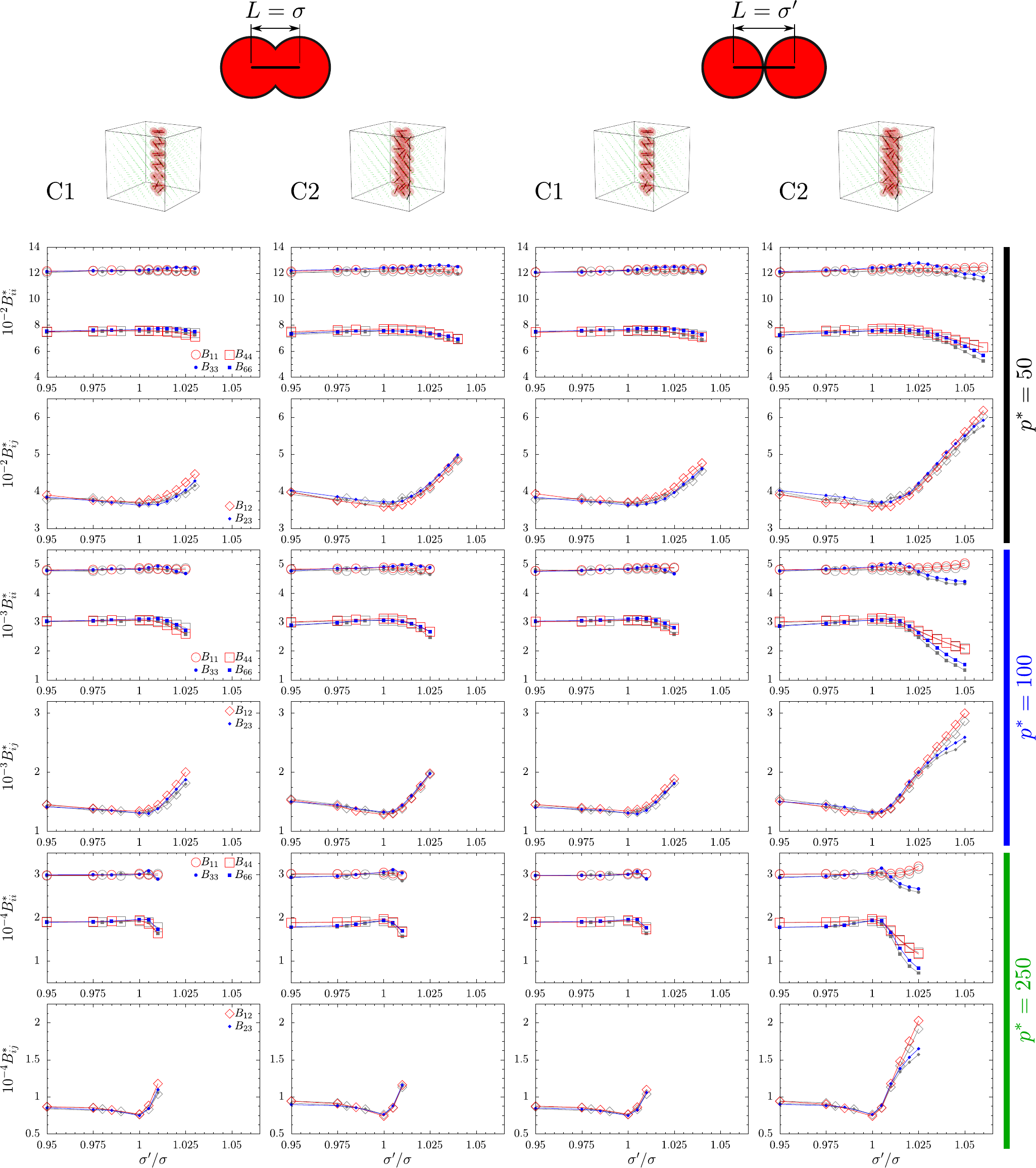}
  \caption{Elastic constants $B_{ij}$ of tetragonal system with $[001]$-nanochannel inclusions filled with hard dimers (in colour) compared with compliances of systems containing $[001]$-nanochannel inclusions filled with hard spheres (in gray). The pressure values for the corresponding data sets have been marked with colour labels on the right. The data has been arranged in columns for systems where dimers had length equal to $\sigma$ (two columns on the left) and for systems where dimers had length equal to $\sigma'$ (right two columns). The columns correspond to the given width of nanochannels, indicated at the top with C$x$ markers.}
  \label{FG_BijBenchmark}
\end{figure*}

\begin{figure*}
  \includegraphics[width=\textwidth]{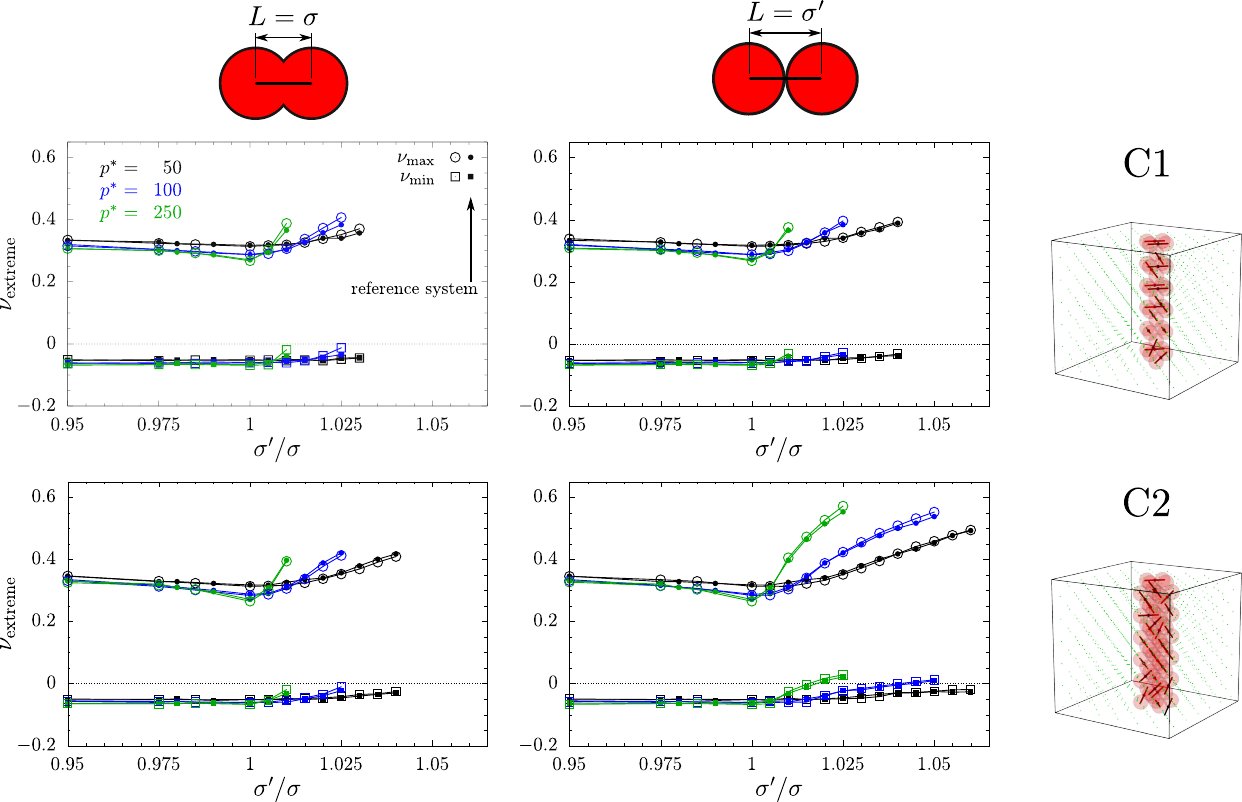}
  \caption{Comparison of extreme values of {\pr} of studied systems. Dimer systems (open symbols) are plotted against {\fcc} crystal containing $[001]$-nanochannel inclusions filled only with spheres (filled symbols). The data has been divided into plots corresponding to systems where dimers had length $L=\sigma$ (left column) and $L=\sigma'$ (right column). Different sizes of nanochannels have been plotted in rows: narrower, C1 nanochannel (top row) and wider, C2 nanochannel (bottom row).}
  \label{FG_NuExBenchmark}
\end{figure*}

Introduction of an array of nanochannels oriented in $[001]$ direction has been proven to lower the symmetry of the crystal from the cubic to tetragonal one~\cite{JwnKwwKvt2019PSSB,JwnKvtKww2022PSSB}, the latter being described by six independent matrix elements. Thus, the elastic compliance matrix has the form:
\begin{equation}
  \matr{S} = \begin{bmatrix}
    S_{11} & S_{12} & S_{23} & 0      & 0      & 0      \\
    \cdot  & S_{11} & S_{23} & 0      & 0      & 0      \\
    \cdot  & \cdot  & S_{33} & 0      & 0      & 0      \\
    \cdot  & \cdot  & \cdot  & S_{44} & 0      & 0      \\
    \cdot  & \cdot  & \cdot  & \cdot  & S_{44} & 0      \\
    \cdot  & \cdot  & \cdot  & \cdot  & \cdot  & S_{66}
  \end{bmatrix}\ .
  \label{EQ_S}
\end{equation}
The $\tS$ matrix is the Voigt representation~\cite{Nye1957CP} of elastic compliance tensor defined in eq.~(\ref{EQ_Sijkl}).

To directly compare the results from this study with models studied earlier, the sizes of inclusions studied here match the ones studied in~\cite{JwnKwwKvt2019PSSB,JwnKvtKww2022PSSB}. Concentration of inclusion particles is $3.47\%$ and $6.25\%$ for C1 and C2 systems respectively. In Figure~\ref{FG_SijBenchmark} the values of elastic compliance matrix elements , see eq.~(\ref{EQ_S}) above, for the studied systems with dimer inclusions (plotted in red and blue) have been plotted against the results for the systems with inclusions of respective concentration but filled only with spheres (plotted in gray colour). The systems have been studied under three values of the reduced pressure $p^*=50, 100, 250$ (indicated in the Figure). To facilitate the comparison, the diagonal and off-diagonal components have been plotted separately. For systems with inclusions composed only of spheres, the value $\s=1$ corresponds to pure {\fcc} crystal of hard spheres (a cubic reference system). However, for systems containing dimer inclusions this is not the case, as even for $\s=1$ dimer bonds introduce additional constrains to the crystal and the structure is missing the 4-fold symmetry axis.

It has been shown that~\cite{JwnKwwKvt2019PSSB} single $[001]$-nanochannel inclusions significantly reduce the value of the {\pr} at high values of $\s$ ($\sim 1.09$). Unfortunately, no stable systems at such high values of $\s$ were obtained when dimers were introduced into nanochannels. Thus, the presented reference results have been limited to the respective ranges of $\s$ obtained for respective studied systems. In Figure~\ref{FG_SijBenchmark} it can be seen that typically the diagonal components of $\tS$ (namely $S_{11}$, $S_{33}$, $S_{44}$, and $S_{66}$) are lower for dimer inclusions, especially for wider (C2) nanochannels and lower pressures. As one would expect, the introduction of additional constraints to the system makes it more rigid and less prone to volume changes. What is interesting is that even such a small number of molecules in studied systems ($15$ for {\cb} and $27$ for {\cc}) can exert noticeable hardening effect over the {\fcc} crystal. In the case of the off-diagonal compliance elements ($S_{12}$, and $S_{23}$) one can see almost no impact of the dimers forming inclusions. Only a slight increase of $S_{23}$ can be observed at the lowest pressure for systems with $L=\sigma'$; for the definition of $L$ see Fig.~\ref{FG_CH_and_MOL}. However, an interesting fact can be observed when examining the behaviour of $S_{23}$ between {\cb} and {\cc} systems at the lowest pressure $p^*=50$. Regardless of whether the dimers have the length $L=\sigma$ or $\sigma'$, the $S_{23}$ values for dimer {\cb} systems are very close to their spherical counterparts for lower values of $\s$ and tend to decrease when the latter increase. Whereas, the {\cc} dimer systems exhibit higher values of $S_{23}$ for the most of the studied values of $\s$, and only match the reference data for the highest obtained values of inclusion sphere diameters. Figure~\ref{FG_BijBenchmark} presents a similar comparison for elastic constants. The latter are obtained by inversion of elastic compliance matrix. In the figure it can be seen that, systematically, values of $B_{33}$ and $B_{66}$ in systems containing dimers are higher than in respective systems with inclusions containing only spheres. In the case of the wider {\cc} nanochannel for $L=\sigma'$, even a slight increase in $B_{11}$ can be noticed for all studied pressures, when $\s>1$.

Figure~\ref{FG_NuExBenchmark} presents the values for the extreme {\pr} for studied systems with dimer nanochannels compared with reference data~\cite{JwnKwwKvt2019PSSB,JwnKvtKww2022PSSB} for systems with respective inclusions filled with spheres. It can be seen that regardless of the dimer lengths $L$ and the sizes of nanochannels the {\pr} for dimer systems (open symbols) closely follow the results for reference systems (smaller, filled symbols). Even when $\s=1$, which does not correspond to a cubic crystal, one observes no noticeable difference in the {\pr}. For higher values of $\s$ at higher pressures the PR of systems with dimers does increase slightly faster than in the reference systems, but the differences are not significant.

\section{Conclusions}
\label{SEC_Conc}

In this work a comparison of elastic properties between different systems containing nanochannel inclusions have been presented. Systems with nanochannel inclusions oriented in $[001]$ direction and filled with simple, rigid, molecules interacting with hard potential have been compared with systems containing inclusions with matching shapes and sizes, but filled only by hard spheres. It has been shown that whereas, the dimer nanochannel inclusions do not provide us with any additional leverage, when considering the changes of {\pr}, over the pure spherical systems, they do have different impact on the values of elastic compliances. It has been demonstrated that introducing a small number of molecules, made of spheres with different diameters, into the {\fcc} crystal of hard spheres one is able to modify its hardness and shear resistance without changing the {\pr} (with respect to the analogous system without additional constrains imposed on the inclusion particles).

\section*{Acknowledgments}
This research was funded in whole or in part by National 
Science Centre, Poland grant No. 2021/05/X/ST3/01200.
The computations were partially performed at Pozna\'n 
Supercomputing and Networking Center (PCSS). The author is much indebted to 
Professor Fabrizio Scarpa for his hospitality in the Department of Aerospace 
Engineering, University of Bristol, UK. For the purpose of Open Access, the 
author has applied a CC-BY public copyright 
licence to any Author Accepted Manuscript (AAM) version arising from this submission.

\end{multicols}

\end{document}